\documentclass[aps,twocolumn,superscriptaddress,floatfix,pra]{revtex4-1}

\usepackage{lipsum}
\usepackage{graphicx}
\usepackage{amsmath,amssymb}
\usepackage{braket}
\usepackage{mathtools}
\usepackage{hyperref}
\usepackage{multirow}
\usepackage{color}
\usepackage[normalem]{ulem}
\usepackage{amsfonts}
\usepackage{float}
\usepackage{pdfpages} % include appendix
\makeatletter
\usepackage{subfigure}

\def\beq{\begin{equation}}
\def\eeq{\end{equation}}
\def\bea{\begin{eqnarray}}
\def\eea{\end{eqnarray}}

\AtBeginDocument{\let\LS@rot\@undefined}
\makeatother

\begin{document}

\title{Entanglement and purification transitions in non-Hermitian quantum mechanics}

\author{Sarang Gopalakrishnan}
\affiliation{Department of Physics and Astronomy, CUNY College of Staten Island, Staten Island, NY 10314;  Physics Program and Initiative for the Theoretical Sciences, The Graduate Center, CUNY, New York, NY 10016, USA}

\author{Michael J. Gullans}
\affiliation{Joint Center for Quantum Information and Computer Science, NIST/University of Maryland, College Park, Maryland 20742 USA}

\begin{abstract}

A quantum system subject to continuous measurement and post-selection evolves according to a non-Hermitian Hamiltonian. We show that, as one increases the rate of post-selection, this non-Hermitian Hamiltonian can undergo a spectral phase transition. On one side of this phase transition (for weak post-selection) an initially mixed density matrix remains mixed at all times, and an initially unentangled state develops volume-law entanglement; on the other side, an arbitrary initial state approaches a unique pure state with low entanglement. We identify this transition with an exceptional point in the spectrum of the non-Hermitian Hamiltonian, at which PT symmetry is spontaneously broken. We characterize the transition as well as the nontrivial steady state that emerges at late times in the mixed phase using exact diagonalization and an approximate, analytically tractable mean-field theory; these methods yield consistent conclusions.  

\end{abstract}
\vspace{1cm}

\maketitle

\section{Introduction}

The dynamics of open quantum systems is a central theme in various branches of contemporary many-body physics~\cite{breuer2002theory, schaller2014open, rotter2015review, diehl2008}. In many circumstances, the environment to which an open system is coupled can be modeled as Markovian. When a system is coupled to a Markovian environment, one can equivalently regard it as undergoing repeated weak measurements at each instant, with each measurement involving a new measuring apparatus. The system and the measuring apparatuses are collectively in a pure state.  For each set of measurement outcomes, the system itself is in a pure state, but if one traces over measurement outcomes the system is in a mixed state that evolves according to a Lindblad master equation~\cite{GKS, Lindblad1976}. A set of measurement outcomes is called a ``quantum trajectory''~\cite{dalibard1992wave}. Recent work, motivated by quantum circuits, has explored the properties of wavefunctions associated with typical individual trajectories. Unexpectedly, these single trajectories undergo a phase transition in their entanglement properties as the rate of measurements is increased~\cite{Li2018, Skinner2019, Chan2019, Li2019, Gullans2019A, Gullans2019B, Fan2020, Choi2019b, Jian2019, Cao2019, Bao2020, Nahum2019, Zabalo2020, Tang2020, Vasseur2020, Li2020, Fisher2020, Lavasani2020, Hsieh2020, Alberton2020, ippoliti2020entanglement, ippoliti2020postselection, fidkowski2020dynamical, nahum2020measurement, iaconis2020measurement}: for sufficiently weak or sparse measurements, the bipartite entanglement of the system along a typical trajectory grows to become a volume law (as it does under purely unitary dynamics); for strong or dense measurements, a typical trajectory has area-law entangled wavefunctions. This ``measurement phase transition'' has also been interpreted in terms of the ability of the dynamics to hide information in nonlocal correlations that the measurements do not probe~\cite{Choi2019b, Gullans2019A, Gullans2019B}. This transition is, however, invisible in the trajectory-averaged evolution of the density matrix under the master equation. 

The measurement transition seems to occur for a broad class of models, but its character depends on the precise ensemble of trajectories. The most studied version of this transition involves weighting trajectories by their probability of occurring, i.e., by the Born probabilities of the measurement outcomes. An alternative ensemble is to fix a particular measurement outcome and post-select on it: i.e., to keep only runs with a particular (and thus generally atypical) measurement history; this has been called the ``forced measurement phase transition''~\cite{nahum2020measurement} (see also~\cite{lee_chan, PhysRevLett.104.160601, biella2020many}). In experimental settings, post-selection might be more practical than sampling over all measurement outcomes (e.g., if measurements are done by fluorescence imaging, outcomes where a particle is observed cause heating, which contaminates the subsequent dynamics). For random circuits, there is once again an entanglement transition in this ensemble, though it is believed to fall into a different universality class than that in the Born-rule ensemble~\cite{nahum2020measurement}. The work so far on both of these measurement transitions has focused on problems in which there is some irreducible temporal randomness, either in the dynamics itself or in the measurement outcomes. This temporal randomness makes it unnatural to address measurement phase transitions from the \emph{spectral} perspective that has proved fruitful for understanding quantum chaos, many-body localization, and related phenomena in closed quantum systems~\cite{mukund, Nandkishore2015}. 

\begin{figure}[!t]
\begin{center}
\includegraphics[width = 0.45\textwidth]{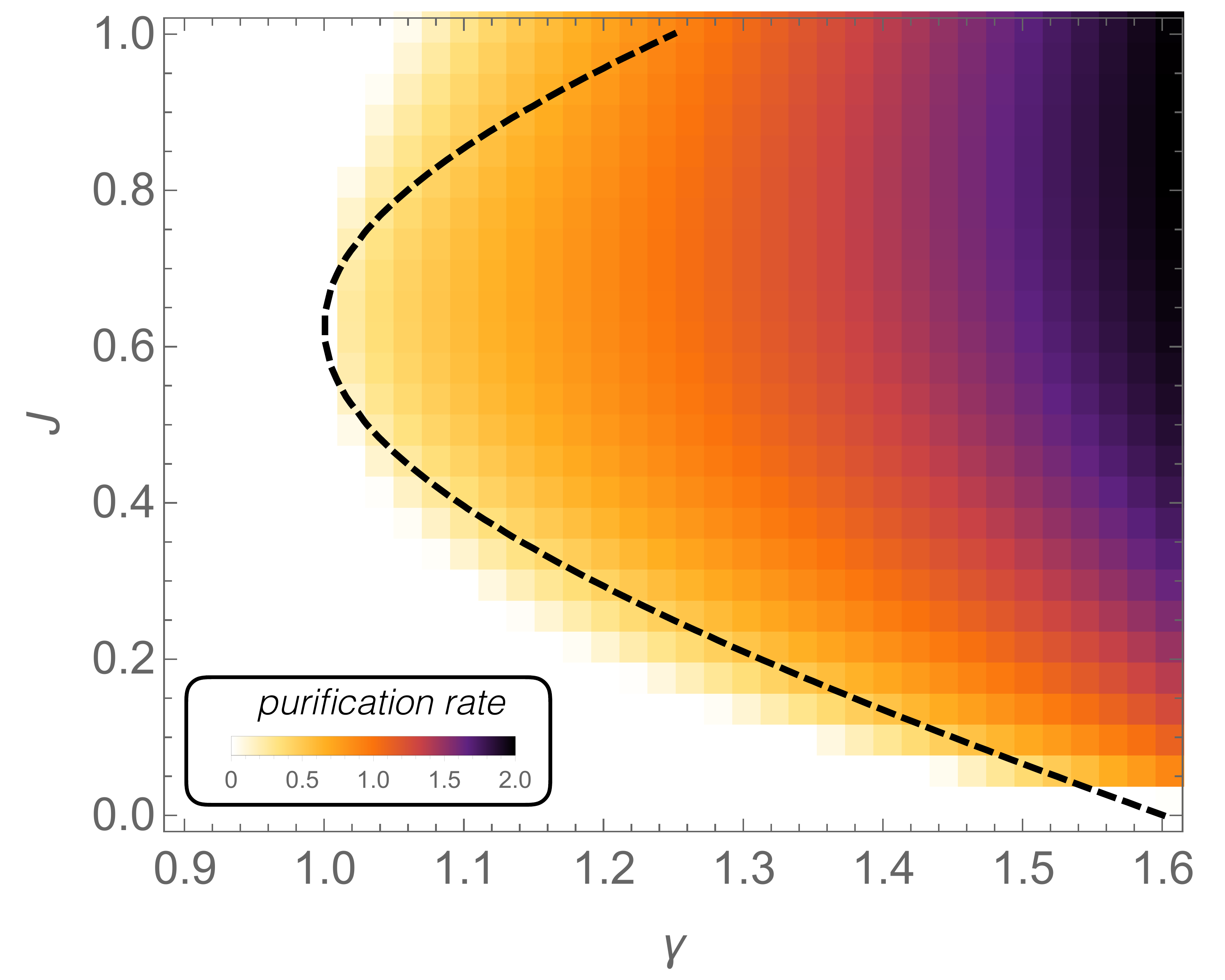}
\includegraphics[width = 0.45\textwidth]{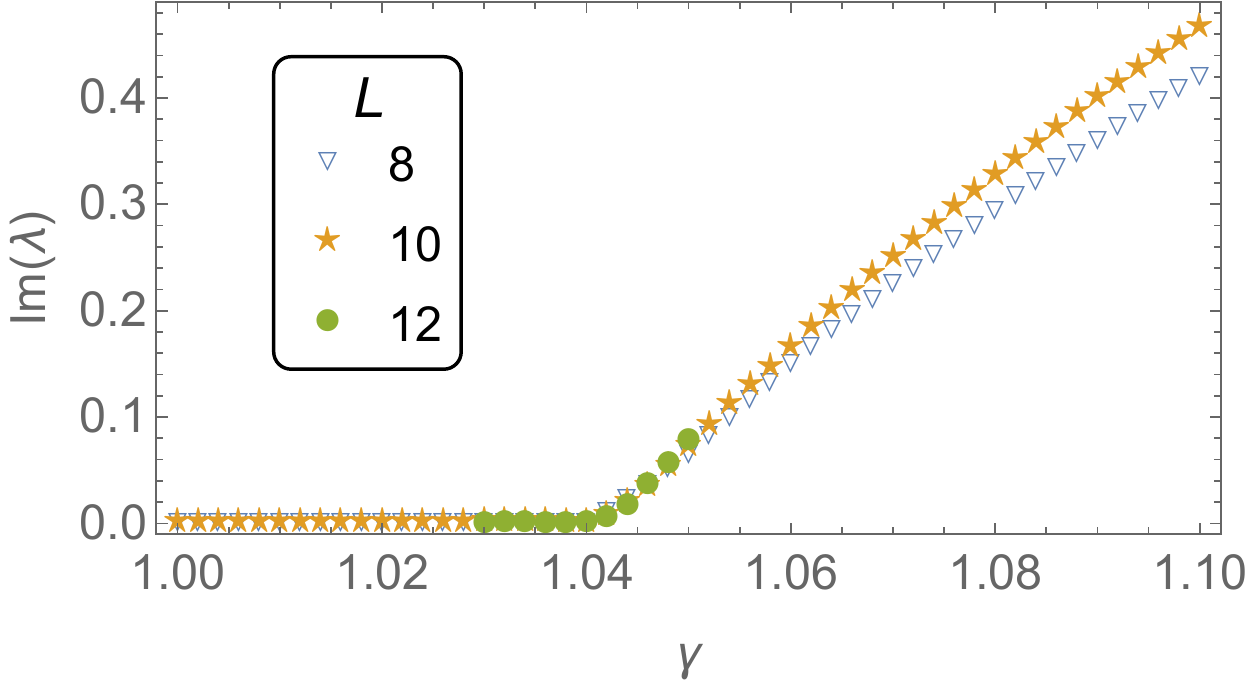}
\caption{Upper panel: Phase diagram of purification rate vs. measurement strength $\gamma$ and interaction strength $J$. The dashed line is the mean-field prediction, which is in good agreement with the numerical data (color map) on one-dimensional spin chains, except at large $J$. Lower panel: size-dependence of the gap opening for $J = 0.95$.}
\label{phasediag}
\end{center}
\end{figure}

The present work aims to fill this gap by studying systems in which the time evolution and the measurement outcomes are fixed: therefore, there is a well-defined non-unitary evolution operator that we can study the spectral properties of. Specifically, we consider a chaotic Ising chain  subject to continuous weak measurements of the Pauli operator $\sigma^y$ on every site. We post-select on the outcome where the weak measurements do not show the spin being in the state $|+y\rangle$. This post-selected dynamics is described by a non-Hermitian Hamiltonian that (apart from a trivial constant) has purely real entries; this non-Hermitian Hamiltonian undergoes a spectral transition. When the measurements are strong, each eigenvalue of the non-Hermitian Hamiltonian has a distinct imaginary part (decay rate), and at late times any initial state gets projected onto the longest-lived eigenstate, which is area-law entangled. However, when the measurements are weaker than a certain threshold, all the eigenvalues of the non-Hermitian Hamiltonian have the \emph{same} decay rate: thus, for example, an initially mixed state remains mixed to arbitrarily late times. (This property is also a defining feature of the volume-law phase in the entanglement transitions studied so far.) Further, an initial product state develops volume-law entanglement, which takes anomalously long to reach its saturated value. Because the Hamiltonian is non-Hermitian, an analogue of this post-selection transition is sharply defined even for few-level systems, and simply corresponds to an exceptional point~\cite{bender1998real} in the spectrum, at which two real eigenvalues merge and then split in the imaginary direction. Unexpectedly, this exceptional point occurs at a nonzero value of dissipation even in the thermodynamic limit, so the mixed phase seems to exist as a true phase in this model. We present numerical evidence for this, and also develop a mean-field theory of the transition that we expect to be valid in high enough dimensions---though it works better than one might expect even in one dimension (Fig.~\ref{phasediag}). 
We then explore the dynamics and the spectral properties of the mixed phase. 
The eigenstates of the non-Hermitian Hamiltonian retain volume-law entanglement throughout the mixed phase, but increasingly violate the eigenstate thermalization hypothesis~\cite{DAlessio2016}. 
A crucial feature of the mixed phase is that even though the eigenvalues of the time-evolution operator are (up to a trivial overall offset) real, the eigenstates are not mutually orthogonal. As the measurement rate increases, these eigenstates get increasingly nonorthogonal, and effectively occupy a much lower-dimensional space, in a way that we will make more precise below. 
This non-orthogonality is responsible for the partial purification of initially fully mixed states.

This work is organized as follows. In Sec.~\ref{background} we introduce background on entanglement and purification transitions as well as non-Hermitian quantum mechanics. In Sec.~\ref{tls} we discuss a toy example: the purification transition for a single qubit. In Sec.~\ref{mf} we extend this to develop a mean-field approximation for the purification transition. In Sec.~\ref{numx} we present numerical evidence from one-dimensional spin chains on the dynamics of initially pure and mixed states across this transition. In Sec.~\ref{lstat} we discuss the spectral statistics and eigenvectors of the non-Hermitian Hamiltonian in the mixed phase. Finally we close with a discussion of the implications of our findings.

\section{Background}\label{background}

\subsection{Quantum channels and trajectories}

A quantum channel, or completely positive trace preserving map~\cite{GKS, nielsen2002quantum}, is a linear map that takes each physical density matrix $\rho$ to another physical density matrix $\mathcal{M}(\rho)$. In general, a quantum channel can be expressed in terms of a set of Kraus operators $K_i$, 
\beq
\mathcal{M}(\rho) = \sum_{i = 1}^{k} K_i \rho K_i^\dagger,
\eeq
such that $\sum_i K^\dagger_i K_i = 1$. (This condition is trivially satisfied, for example, when $K_i$ are projectors onto measurement outcomes.) One can regard Kraus operators as the operators in the system that couple to the external environment (e.g., the measuring apparatus). Markovian time evolution consists of the repeated application of quantum channels, and one can expand the final density matrix in terms of Kraus operators as
\beq
\rho(t) = \sum\nolimits_{i_1, i_2 \ldots i_n} K_{i_n} \ldots K_{i_1} \rho_0 K_{i_1}^\dagger \ldots K_{i_n}^\dagger.
\eeq
Each sequence $\{ K_{i_n} \}$ appearing in this sum is called a ``quantum trajectory.'' When the initial density matrix is a pure state $|\psi\rangle\langle \psi|$, each individual quantum trajectory gives rise to a particular pure state (which would result if one observed precisely that set of measurement outcomes). In the expression above, if we take the initial state to be pure, a particular trajectory has weight $\Vert K_{i_n} \ldots K_{i_1} |\psi\rangle \Vert^2$; for projective measurements this yields the Born probability of the set of measurement outcomes indexed by the Kraus operators along that trajectory.

The continuum limit of a quantum channel is the Lindblad master equation~\cite{GKS, Lindblad1976},
\beq
\dot{\rho} = - i [H, \rho] + \sum_{i = 1}^m \gamma_i \left[ O_i \rho O_i^\dagger - \frac{1}{2} (O_i^\dagger O_i \rho + \rho O_i^\dagger O_i) \right].
\eeq
Here, the $O_i$ are conventionally referred to as jump operators. This is a special limit of a quantum channel in which $k - 1$ of the Kraus operators are proportional to $O_i$, and account for the system interacting with its environment, while the last Kraus operator---which gives rise to the last two terms in the master equation---depletes the norm of the state in the event that it did not get measured~\cite{minev2019catch}. 

In Lindblad time evolution there is therefore a special trajectory along which the detectors never record anything. Unraveling along this special trajectory gives rise to time evolution that is described by an effective non-Hermitian Hamiltonian, $H_{\mathrm{eff}} = H + i \frac{\gamma}{2} \sum_{i = 1}^m O^\dagger_i O_i$. This work is concerned with dynamics under this special trajectory, which can be accessed by post-selection. (Note that there has been some prior work on this topic, although mostly for free-fermion models~\cite{lee_chan, biella2020many}.)

\subsection{Entanglement and purification transitions}

Entanglement transitions have mostly been discussed in the context of qubits (or generally qudits) subject to spatio-temporally random unitary gates and single-site measurements. Variants in which the unitary gates are not random (but the measurement outcomes remain random), or in which unitary evolution is traded off for multiple-site measurements, have also been considered, but do not seem to change the properties of the transition. Most papers on the entanglement transition have considered an ensemble of states in which the probability that a particular trajectory occurs is simply its Born probability. However, a recent work also introduced the case of ``forced measurements'' in which one post-selects on a particular measurement outcome: this is closer in spirit to the problem we are considering here~\cite{nahum2020measurement}. In that work, the forced measurement transition was considered in temporally random circuits; here, we eliminate the temporal randomness. 

The entanglement transition is diagnosed by the entanglement properties of an initial pure state that is evolved to very late times. In the absence of measurements, one expects this state to evolve to a nearly random state with volume-law entanglement entropy for any spatial region. When measurements are present at some subcritical rate $p < p_c$ (where $p_c$ depends on the precise model) the late-time wavefunction is volume-law entangled. However, for $p > p_c$, measurements ``disentangle'' the quantum state so its bipartite entanglement settles to an area law. That a stable volume-law phase exists at all is counterintuitive; one way of understanding it is that the time evolution hides the quantum information present in the initial state in nonlocal correlations that are not affected by the measurement~\cite{Choi2019b}.

A useful alternative way of thinking about these entanglement transitions is as \emph{purification transitions}~\cite{Gullans2019A}. According to this perspective, one evolves an initially fully mixed state under a trajectory $T$, leading to an (un-normalized) density matrix $\rho \propto T T^\dagger$. In the limit of frequent measurements, each measurement lowers the rank of the density matrix, which loses rank at a fixed rate and becomes a pure state on a timescale that is sublinear in system size. In the opposite limit, measurements cannot successfully extract quantum information from the system, so the mixed state remains mixed out to exponentially long times in system size. Thus, if one takes the thermodynamic limit and measures the purity of the density matrix at a time $t \propto L^\alpha$, ($\alpha > 1$ in one dimensional models), one finds a sharp transition between mixed and pure phases. In the models studied so far, the purification and entanglement transitions coincide; we will find that this is also the case here.

\subsection{PT symmetry}

Hermitian matrices have real spectra, but the converse is false: a broader class of matrices, called PT-symmetric matrices~\cite{bender1998real, moiseyev2011non, takasu2020pt, ashida2020non}, share this property. A simple $2 \times 2$ example is the matrix
\beq\label{m0}
M_0 \equiv \left(
\begin{array}{cc}
0 & 1 + b \\ 1 - b & 0
\end{array}
\right),
\eeq
where $b$ is real. The characteristic equation of this matrix is $\lambda^2 = 1 - b^2$, so when $|b| < 1$ the eigenvalues are both real. The eigenvalues collide when $b = 1$, leading to an ``exceptional point.'' (In this simple example the PT symmetry is just the fact that the matrix is purely real, so its eigenvalues must come in complex conjugate pairs, and therefore have to collide before they can wander into the complex plane. However, the phenomenon has more general manifestations as well.) When $b > 1$ both eigenvalues are pure imaginary.

Although PT-symmetric non-Hermitian matrices have real spectra, they lack other features of Hermitian matrices, notably the orthogonality of eigenvectors. In the present example, some algebra yields that the (standard) inner product between two eigenvectors is simply $|b|$. When $b = 1 - \epsilon$, the eigenvectors take the form $|\pm \rangle \approx (1, \pm \sqrt{\epsilon/2})$. At the exceptional point $b = 1$ the eigenvectors become parallel, so the matrix ceases to be diagonalizable. For small $\epsilon > 0$, the two eigenvectors are linearly independent and thus span the entire space. However, the expression for a generic vector in this basis (for example $(0,1) = 1/\sqrt{\epsilon}(|+\rangle - |-\rangle)$) involves large coefficients and approximate cancellations. The matrix that implements the basis transformation becomes singular at the exceptional point, and is ill-conditioned near it. 

Exceptional points are a ``phase transition'' of sorts, but unlike conventional thermal or quantum phase transitions, it does not directly involve a thermodynamic limit. (Physically, of course, systems that are described by non-Hermitian Hamiltonians implicitly involve a classical environment, so there is no rule against such phase transitions.)

\section{Purification of a two-level system}\label{tls}

We now consider the simplest purification problem, which is that of a two-level system. Consider a Lindblad master equation with a single jump operator proportional to $(1 + \sigma^y)$, i.e., a weak measurement in the $y$ basis. The corresponding non-Hermitian Hamiltonian is $H_{\mathrm{eff}} = H_0 + i b (1 + \sigma^y)$. We choose our coordinate system so that $H_0 = \sigma^x$. It follows that $H_{\mathrm{eff}} = M_0 + i b$ where $M_0$ is defined in the toy example~\eqref{m0} above. 

We consider the dynamics of initially mixed states evolving under this non-Hermitian Hamiltonian, $T = \exp(-i H_{\mathrm{eff}} t)$. For any matrix one can write $T = P D P^{-1}$, where the columns of $P$ are eigenvectors of $H_{\mathrm{eff}}$, and $D$ is a diagonal matrix with entries $e^{-i \lambda_k t}$ where $\lambda_k$ are the eigenvalues of $H_{\mathrm{eff}}$. We write this expression out in general, using a coordinate system derived from some reference orthonormal basis (e.g., the computational basis):
\beq\label{tev0}
\rho \propto [(P^\dagger)^{-1}]_{ij} e^{i \lambda^*_j t} (P^\dagger)_{jk} P_{kl} e^{-i \lambda_l t} (P^{-1})_{lm}.
\eeq
Eq.~\eqref{tev0} supports two kinds of behavior, which are illustrated with respect to the toy model. For $|b| > 1$, the eigenvalues are $\lambda_\pm = -i (b \pm \sqrt{b^2 - 1})$. Time evolution therefore causes the two eigenvectors to decay at different rates; at late times, $T$ is (again up to normalization) a projector onto the slowest-decaying eigenvector. In this regime any initial state eventually morphs into the slowest-decaying eigenvector. However, for $|b| < 1$, the eigenvalues decay at the same rate---$\lambda_\pm = -ib \pm \sqrt{1 - b^2}$---so a mixed state remains mixed. The phase factors in Eq.~\eqref{tev0} oscillate, and if one averages over these oscillations (by going into an appropriate ``diagonal ensemble''~\cite{Nandkishore2015}) one finds the time-averaged density matrix
\beq\label{diag}
\rho_{\mathrm{ss}} \propto [(P^\dagger)^{-1}]_{ij} (P^\dagger)_{jk} P_{kj} (P^{-1})_{jm}.
\eeq
Defining the steady-state purity as $\Pi \equiv \mathrm{Tr}(\rho_{\mathrm{ss}}^2)/(\mathrm{Tr} \rho_{\mathrm{ss}})^2$, a straightforward if tedious calculation yields 
\beq
\Pi = \frac{1}{2}(1 + b^2). 
\eeq
Putting this together with the $|b| > 1$ regime, the steady-state purity in this case is a continuous function of $b$, with a cusp at the exceptional point. 

\begin{figure}[tb]
\begin{center}
\includegraphics[width = 0.45\textwidth]{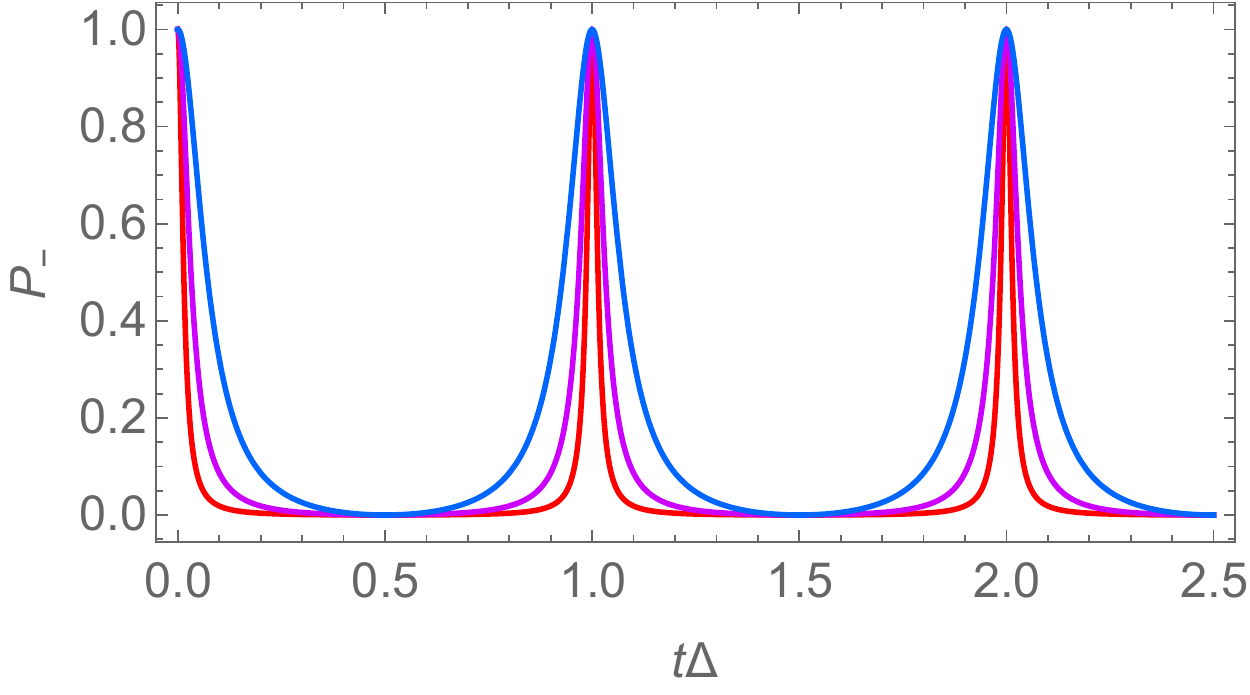}
\caption{Dynamics of the state $|- \rangle \equiv (0,1)$ under non-Hermitian dynamics near the exceptional point. The $x$ axis is time rescaled by the eigenvalue difference $\Delta$ of the TLS Hamiltonian~\eqref{m0}; the $y$ axis is the probability of measuring the time-dependent state to be in the state $| - \rangle$. The three curves are for $\Delta = 0.1, 0.02, 0.004$ respectively; the revivals get narrower as one approaches the exceptional point.}
\label{tlsfig}
\end{center}
\end{figure}

It is instructive to consider dynamics precisely at the exceptional point. Here, the Hamiltonian has only one eigenvector, $(1,0)$. A generic initial state $|\psi\rangle = (\cos \theta, \sin \theta)$ evolves to the (unnormalized) state
\beq
|\psi(t) \rangle = (\cos \theta + i t \sin \theta, \sin \theta).
\eeq
Thus, any initial state eventually points along the eigenvector; the perpendicular components are relatively suppressed as $1/t$. Accordingly, any initial state purifies as
\beq
1 - \Pi(t) \sim 1/t^4.
\eeq
That a ``critical exponent'' can exist for a two-level system is one of the peculiarities of non-Hermitian quantum mechanics. Slightly on the mixed side of the phase transition, the purity oscillates as in Fig.~\ref{tlsfig}; the decay of the component transverse to the eigenvector is interrupted by periodic revivals, set by the period $1/[2\sqrt{2(1-b)}]$. These become increasingly narrow as one approaches the exceptional point.

Finally, we find that the expectation value 
\beq\label{expval}
\langle \sigma^z \rangle \equiv \mathrm{Tr}(\sigma^z \rho_{\mathrm{ss}}) / \mathrm{Tr} \rho_{\mathrm{ss}} = -|b|.
\eeq 

\section{Mean-field theory}\label{mf}

The analysis above can directly be extended to a simple Weiss-type mean-field theory for purification transitions in high-dimensional systems. Consider a non-Hermitian Hamiltonian of the form
\beq
H_{\mathrm{eff}} = \sum_i \left[ h_i \sigma^z_i + g_i \sigma^x_i + i \gamma (1 + \sigma^y_i) \right] + \sum_{ ij} J_{ij} \sigma^z_i \sigma^z_j,
\eeq
which we will explore numerically in what follows. For now we treat the connectivity $J_{ij}$ as general. One can decouple this interaction to get a Hamiltonian for a single spin subject to a complex field, 
\beq
h_{\mathrm{eff}}( \langle \sigma^z \rangle) = \left(h_i + \sum\nolimits_j J_{ij} \langle \sigma^z_j \rangle\right) \sigma^z + g_i \sigma^x + i \gamma (1 + \sigma^y).
\eeq
This decoupling is valid in the high-dimensional limit, and more generally is a reasonable variational estimate for $\gamma$ small or large (since in both limits the steady-state density matrix is unentangled). The mean-field equation can then be solved self-consistently for $\langle \sigma^z \rangle$ following the arguments in the previous section. This yields the self-consistency condition
\beq\label{scmf}
\langle \sigma^z_i \rangle = -\frac{g_i \gamma_i}{g_i^2 + \left( h_i - \sum_j J_{ij} \langle \sigma^z_j \rangle \right)}.
\eeq
Assuming a translation-invariant state, this reduces to a cubic equation for $\langle \sigma^z \rangle$. For a translation-invariant system, this mean-field theory predicts that the phase transition occurs when 
\beq\label{mfboundary}
g \gamma = g^2 + (h - J z)^2, 
\eeq
where $z$ is the coordination number of each site. This mean-field theory agrees well with numerics even for one-dimensional spin chains, though there are discrepancies at interaction strengths $J \approx 1$ (Fig.~\ref{phasediag}).

\section{One-dimensional spin chains}\label{numx}

We now turn to numerical results on spin chains evolving under the following non-Hermitian Hamiltonian:
\beq\label{scham}
H = \sum_i \left[ h_i \sigma^z_i + \sigma^x_i + i \gamma (1 + \sigma^y_i) + J \sigma^z_i \sigma^z_{i+1} \right].
\eeq
We will explore both the translation-invariant model and a model with weak randomness in the longitudinal fields $h_i \in [h_0 - \epsilon, h_0 + \epsilon]$ where $\epsilon \leq h_0 / 10$. We will further restrict ourselves to $h_0 = 1.25$---it seems that $h_0 = 1.25, J = 0.95$ gives the cleanest chaotic level statistics in the Hermitian limit, which will be important in the next section. (We remark that this model has often been used as a ``generic'' chaotic model, see e.g.~\cite{kim2013ballistic}.)

The Hamiltonian~\eqref{scham} has a special line at $\gamma = 1$. Along this line, $\sigma^x_i$ and $\sigma^y_i$ combine to form the raising operator $\sigma^+_i$. Consequently, $H$ is an upper triangular matrix and its eigenvalues are precisely its diagonal entries, i.e., they coincide with the eigenvalues of $H' = i \gamma L + \sum_i (\pm h_i \sigma^z_i + J \sigma^z_i \sigma^z_{i+1})$. Since these diagonal entries are real (up to the overall offset), this solvable line demonstrably lies in the mixed phase, consistent with our numerics and mean-field theory. 

The phase diagram of the clean model is plotted in Fig.~\ref{phasediag}. The extent of the mixed phase is non-monotonic, with purification happening soonest when $J \approx h_0 / 2$. This matches the mean-field prediction: at $J = h_0/2$, the applied longitudinal field exactly cancels the self-generated mean field at the transition. Finite-size effects across the phase diagram are very weak (lower panel of Fig.~\ref{phasediag}). As one would expect from the example of the two-level system, the purification rate rises linearly away from the transition. 

\begin{figure*}[tb]
\begin{center}
\includegraphics[width = 0.32\textwidth]{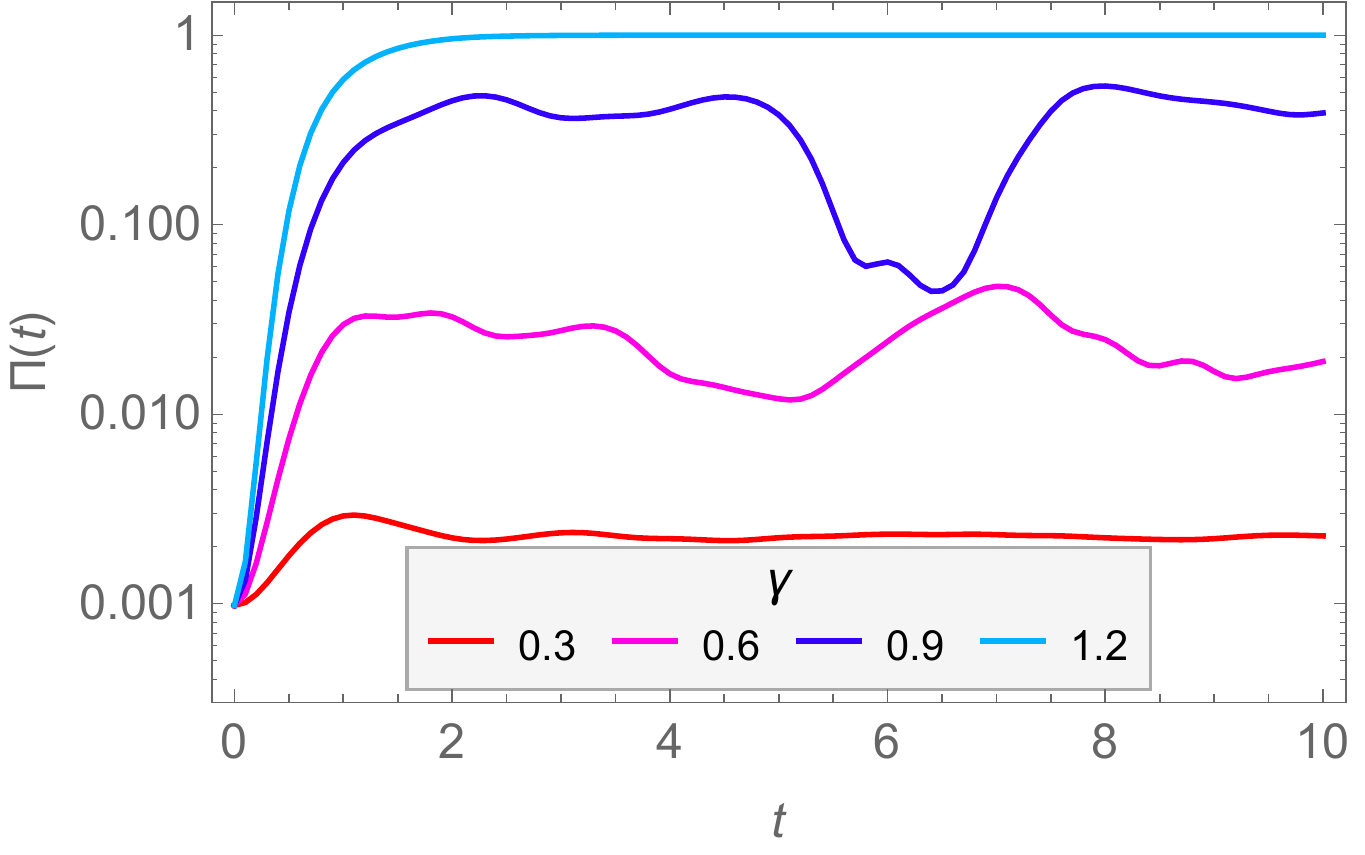}
\includegraphics[width = 0.31\textwidth]{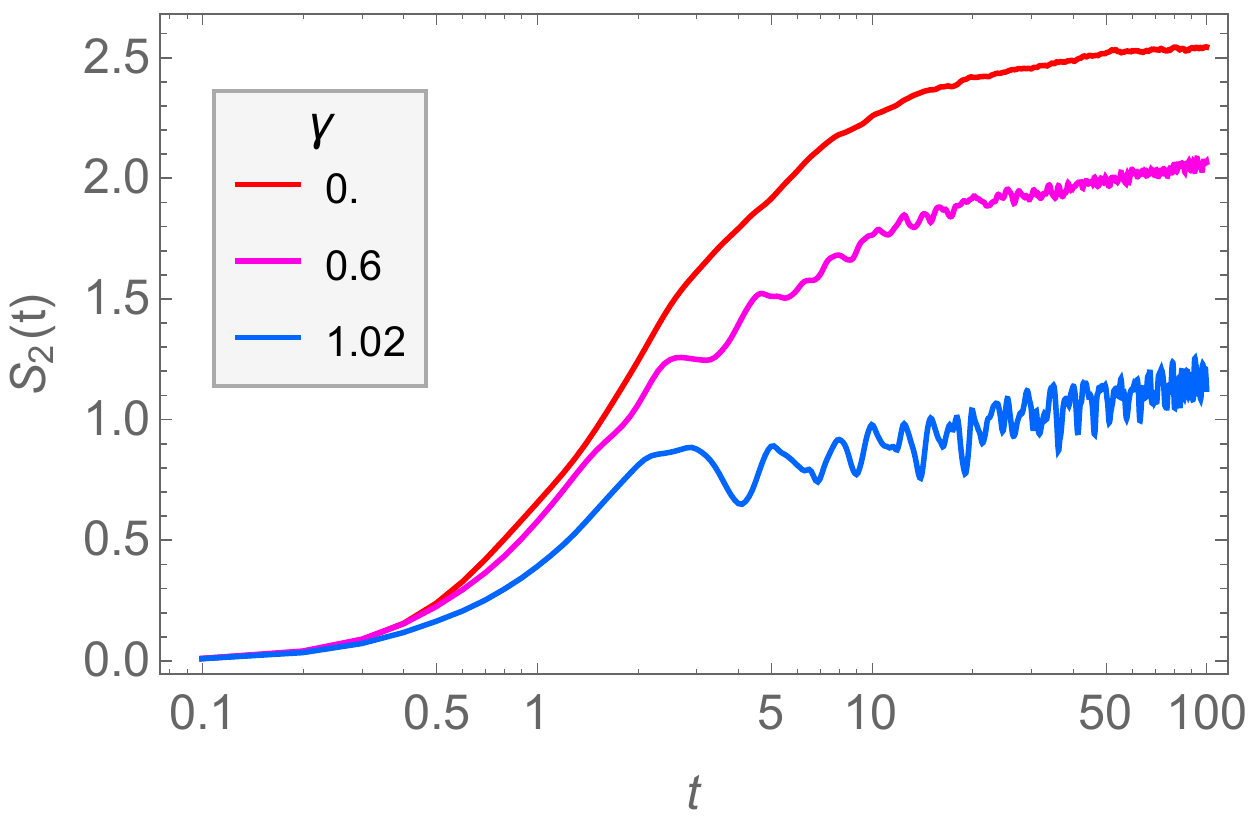}
\includegraphics[width = 0.31\textwidth]{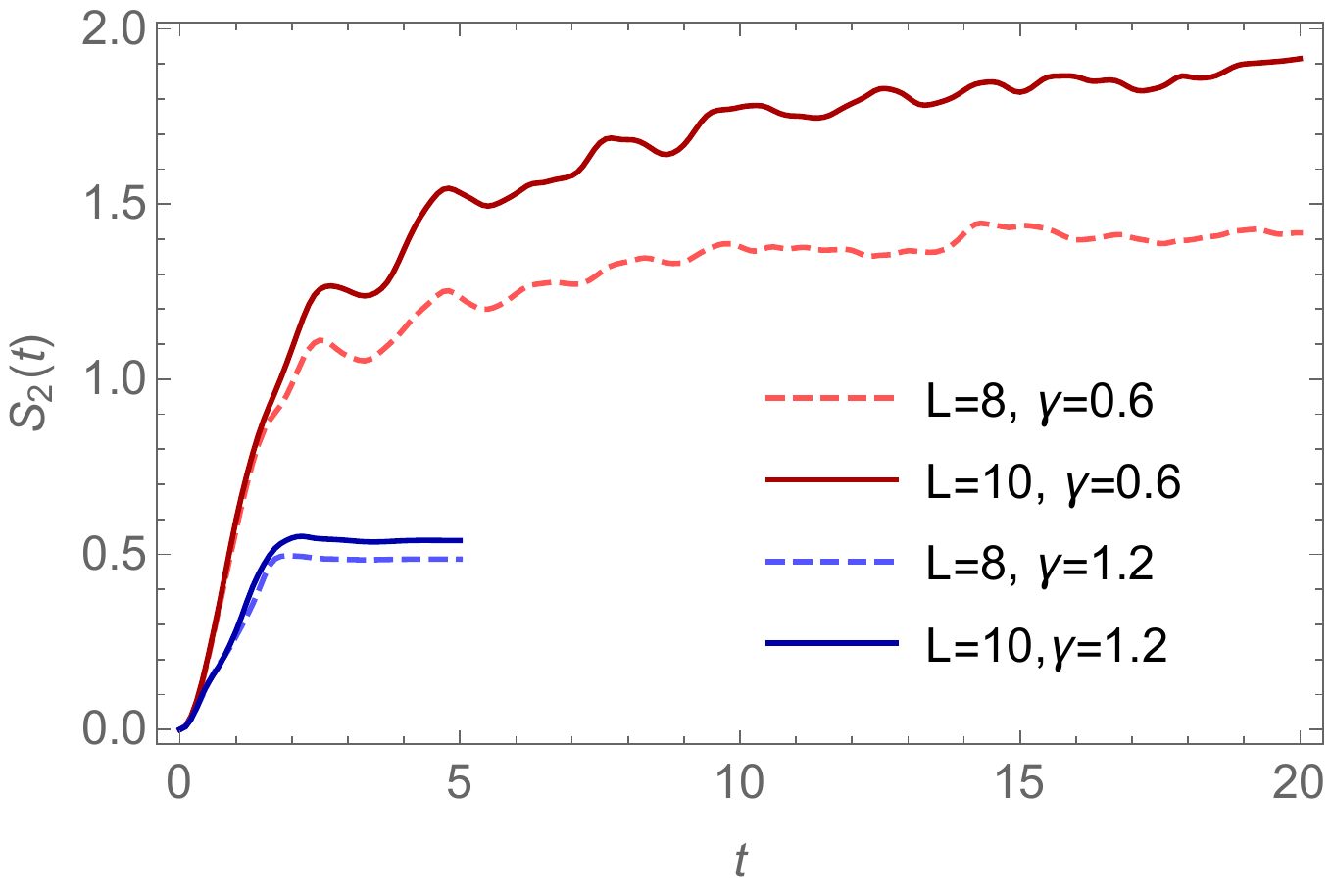}
\caption{Left panel: Evolution of the purity of an initially maximally mixed state, for various values of $\gamma$ and for $L = 10$. The purity undergoes increasingly slow oscillations. Central panel: evolution of the half-cut second R\'enyi entropy of a product state in the mixed phase, vs. $\gamma$, for $L = 10$; a quick initial growth is followed by a much slower, possibly logarithmic, growth. Right panel: size-dependence of $S_2$ in the pure and mixed phases. In the pure phase one sees rapid saturation to a roughly size-independent value, whereas in the mixed phase the entanglement exhibits slow growth and saturates to a volume law with a $\gamma$-dependent prefactor.}
\label{dynamics}
\end{center}
\end{figure*}

We now turn to the dynamics of entanglement and purity in the two phases (Fig.~\ref{dynamics}). Starting from the maximally mixed state, the instantaneous purity $\Pi(t)$ saturates exponentially to unity in the pure phase, as expected. In the mixed phase, it undergoes persistent oscillations around a mean value that is much larger than that of a fully mixed state. (One might expect the purity to be monotonic; while this is true \emph{on average}, it is not true in individual trajectories, as also recently observed in Ref.~\cite{fidkowski2020dynamical}). Both the amplitude and period of these oscillations grow as $\gamma$ is increased. If one starts instead with a random pure product state, the entanglement once again saturates to a small, roughly size-independent value in the pure phase, but grows slowly in the mixed phase. The time $S_2$ takes to saturate is surprisingly long; the origin of this long timescale is an interesting topic for future work.

\begin{figure}[tb]
\begin{center}
\includegraphics[width = 0.45\textwidth]{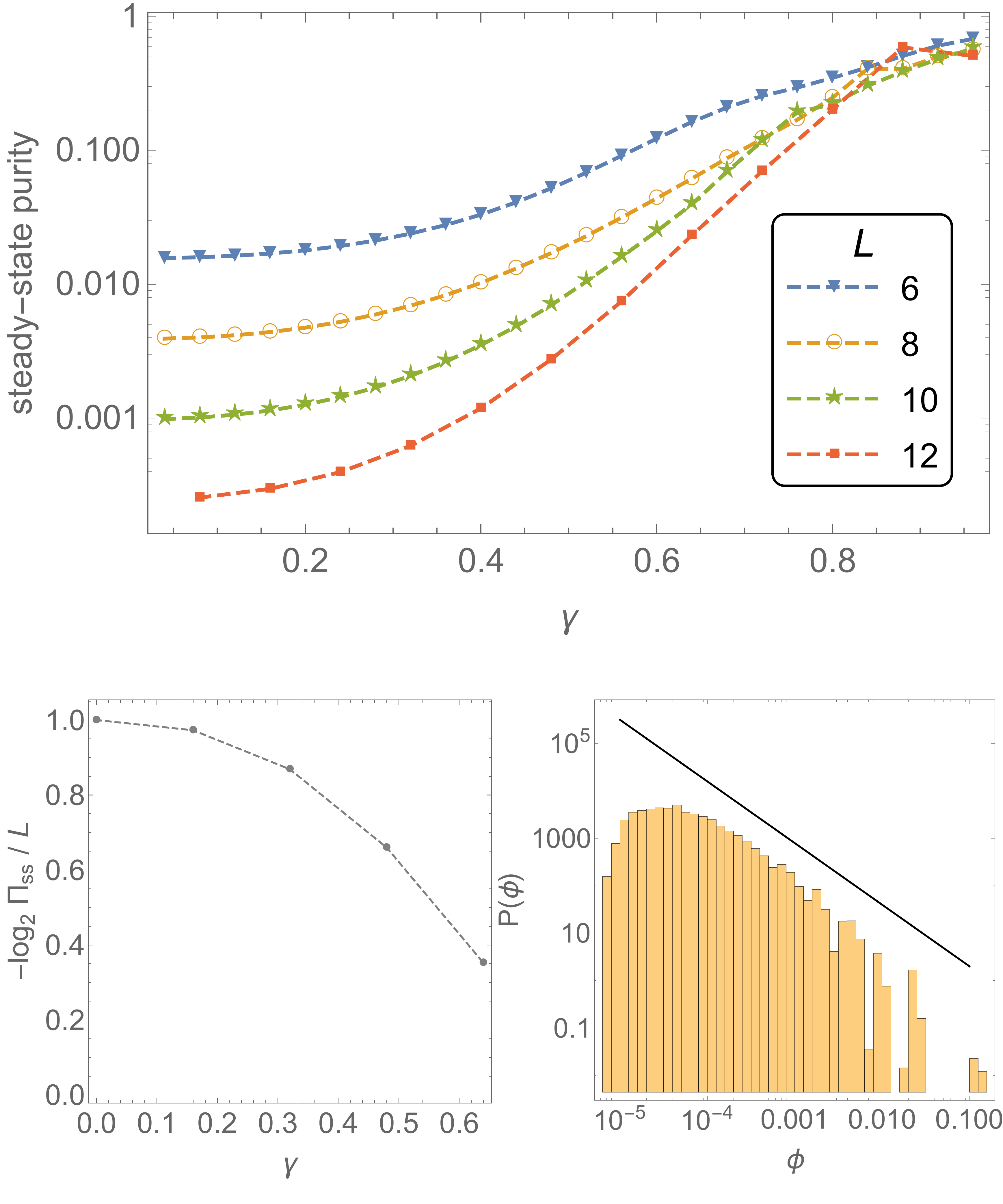}
\caption{Properties of the time-averaged density matrix $\rho_{\mathrm{ss}}$. Upper panel shows the purity vs. $\gamma$; the data for $\gamma \agt 0.7$ are noisy so the trends are inconclusive. The purity goes as $2^{-cL}$, with a coefficient $c$ that decreases with $\gamma$ (lower left). Lower right: histogram of the eigenvalues $\phi$ of $\rho_{\mathrm{ss}}$ for $\gamma = 0.6$ and $L=12$; there appears to be a power-law tail out to large eigenvalues, with a nonuniversal exponent.}
\label{diagdm}
\end{center}
\end{figure}

The slow oscillations in the instantaneous purity make it hard to extract a saturation value; we instead consider the time-averaged density matrix $\rho_{\mathrm{ss}}$ [Eq.~\eqref{diag}]. As one would expect from the dynamics, the purity of $\rho_{\mathrm{ss}}$ increases smoothly with $\gamma$. In general, the purity goes as $\Pi_{\mathrm{ss}} \sim 2^{-c L}$, where $c$ decreases from unity (in the unitary limit) to some much smaller value (Fig.~\ref{diagdm}). This corresponds to the second R\'enyi entropy of $\rho_{\mathrm{ss}}$ following a volume law with a continuously varying prefactor, just as it does in the standard purification transition. Note that $\Pi_{\mathrm{ss}}$ in the clean system evolves non-monotonically and chaotically with $\gamma$. To get a smooth curve, we introduce $1 \%$ disorder in the couplings and average over this disorder (equivalently, one could imagine performing a moving average over $\gamma$). For $\gamma \agt 0.7$, with the amount of averaging we have performed, the purity shows no clear trend with system size; in this regime, the transformation into the eigenbasis is ill-conditioned, leading to potential numerical instabilities. Finally, we note that the eigenvalue spectrum of $\rho_{\mathrm{ss}}$ exhibits a power-law tail, with an apparently nonuniversal exponent. The existence of this tail suggests that different R\'enyi entropies of $\rho_{\mathrm{ss}}$ will have distinct volume-law coefficients. 

\section{Spectral properties}\label{lstat}

\begin{figure}[tb]
\begin{center}
\includegraphics[width = 0.45\textwidth]{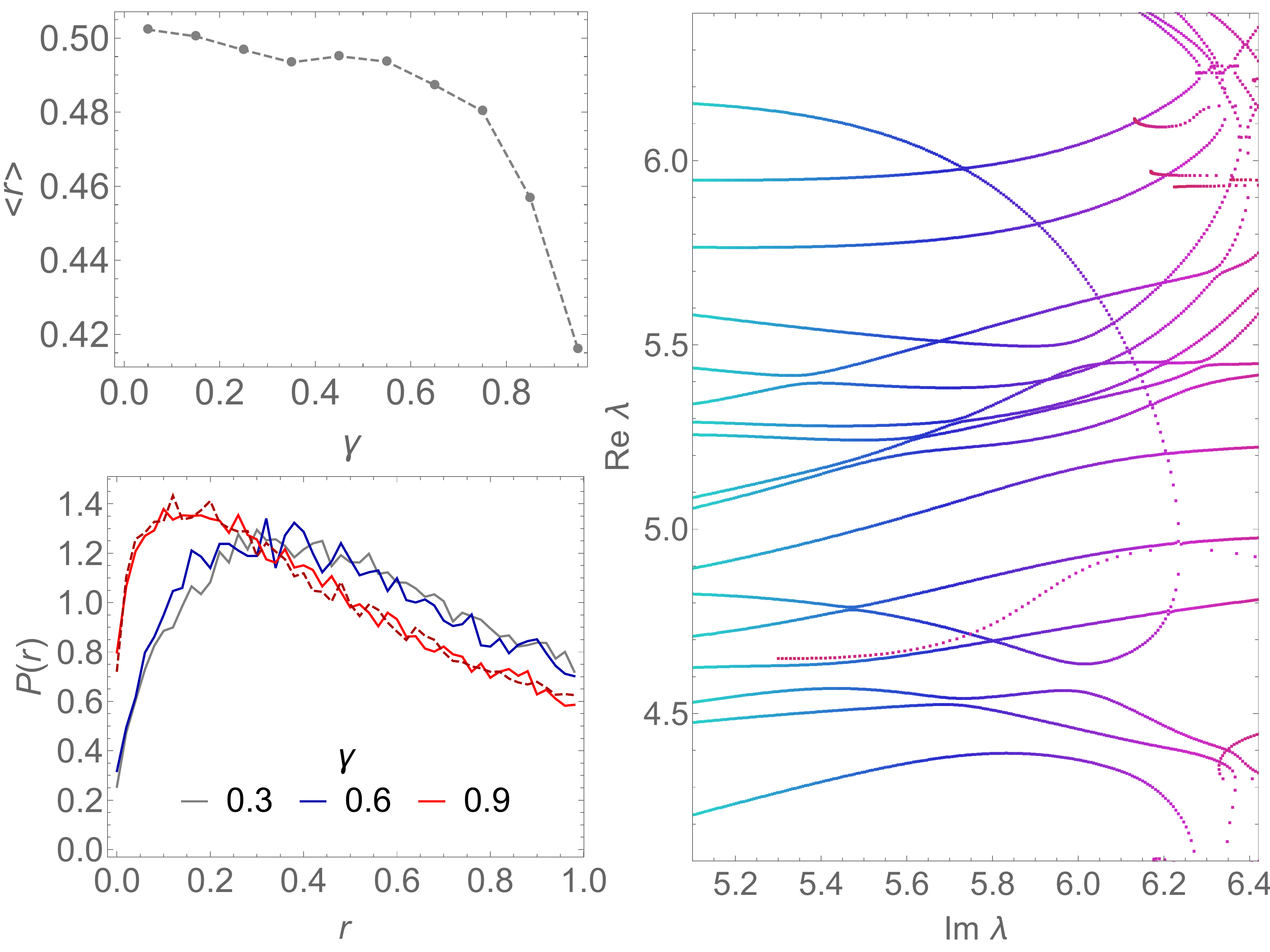}
\caption{Level statistics of the non-Hermitian Hamiltonian throughout the mixed phase. The longitudinal field is subject to $10\%$ disorder in order to break any spatial symmetries. The level spacing ratio $r$ shows clear signs of level repulsion both in its average (upper left) and distribution (lower left), and the level statistics near the transition are intermediate. The two red curves are for $L = 10, 12$ respectively and show good convergence. Right panel follows the spectral ``dynamics'' for a particular disorder realization as $\gamma$ is increased (i.e., as one moves to the right along the figure). There are clear signs of level repulsion (anticrossings), but there also appear to be some near-crossings that are not symmetry-protected..}
\label{levelstats}
\end{center}
\end{figure}

\begin{figure}[tb]
\begin{center}
\includegraphics[width = 0.45\textwidth]{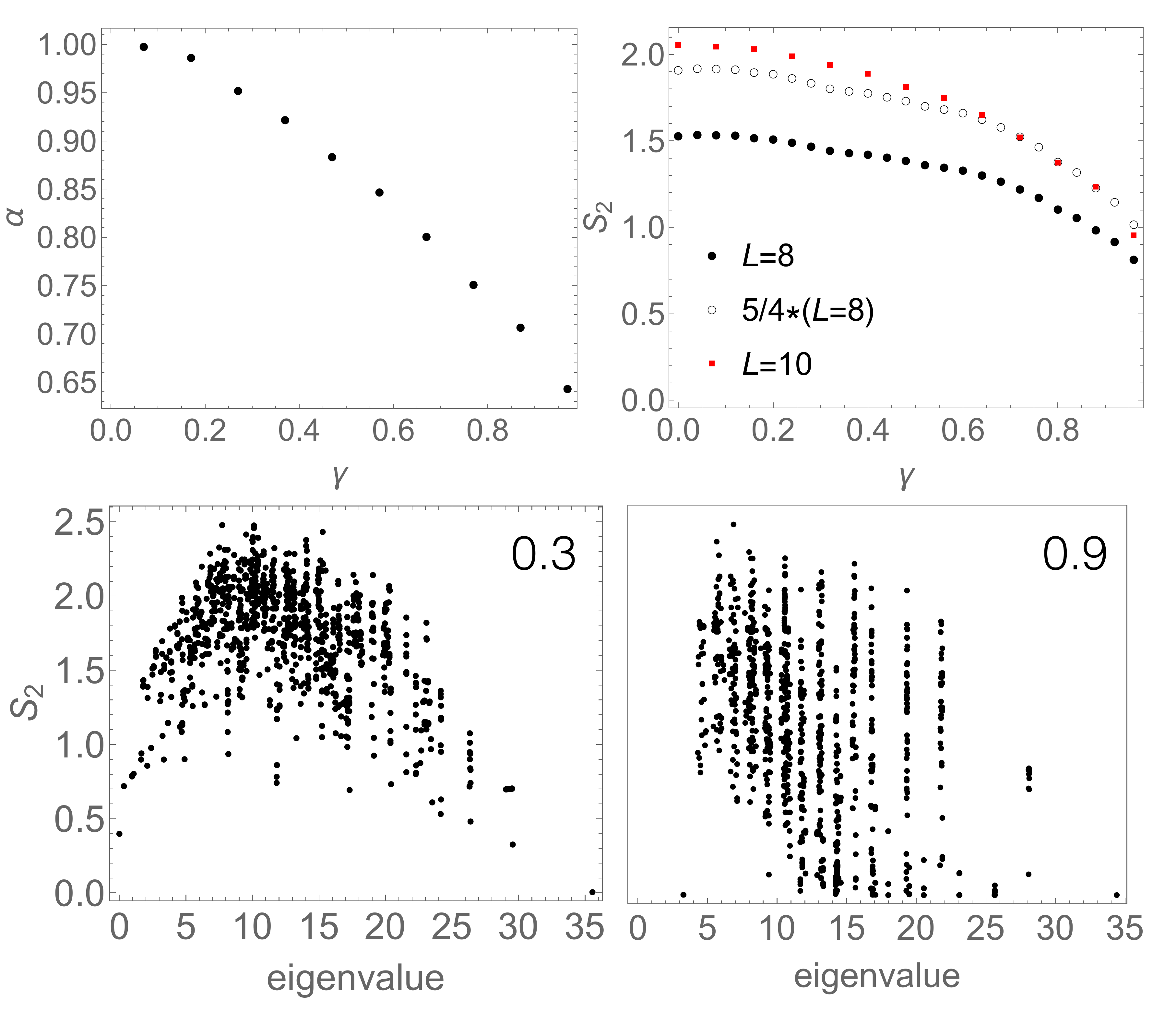}
\caption{Upper left: The effective dimension of the space spanned by the eigenvectors of the non-Hermitian Hamiltonian is $2^{\alpha L}$; upper panel plots $\alpha$ vs. $\gamma$. Upper right: mean eigenstate entanglement vs. $\gamma$; our results are consistent with a volume law. The lower panel illustrates how eigenstate thermalization fails as $\gamma$ is increased, by plotting the second R\'enyi entropy vs. eigenvalue for two different values of $\gamma$.}
\label{eigvecs}
\end{center}
\end{figure}

The previous section summarized the evidence that the non-Hermitian Hamiltonian undergoes a purification and entanglement transition that shares many features with the transition seen in random circuits. In the present context this transition can also be analyzed in terms of the spectral structure of the time-evolution operator (or equivalently of the non-Hermitian Hamiltonian). Accordingly we now discuss, in turn, the properties of eigenvalues and eigenvectors. 

The eigenvalue statistics is illustrated in Fig.~\ref{levelstats}. Following now-standard practice~\cite{Oganesyan2007} we characterize these via the level-statistics ratio $r \equiv \mathrm{min}(|\lambda_i - \lambda_{i-1}|, |\lambda_{i+1} - \lambda_i|) / \mathrm{max} (|\lambda_i - \lambda_{i-1}|, |\lambda_{i+1} - \lambda_i|)$. Recall that in the mixed phase, all eigenvalues have the same imaginary part so eigenvalue differences are purely real. For a chaotic real Hamiltonian, $\langle r \rangle \approx 0.53$ while for a localized or integrable Hamiltonian, we expect $\langle r \rangle \approx 0.39$ (Poisson level statistics). Even for moderately large $\gamma$, the $r$ ratio is very close to its value in the Hermitian limit. Nearer to the transition, however, level repulsion appears to get weaker, and the level spacing ratio dips toward the Poisson value. This dip does not appear to be a finite-size effect: the histogram of $r$ appears well converged with respect to system size. The persistence of level repulsion, and the origin of the spectral transition, can be seen in the right panel of Fig.~\ref{levelstats}: two levels come together and collide at an exceptional point, after which they wander into the complex plane. The complicated balance between level repulsion (which stabilizes the mixed phase) and level attraction (which eventually kills it) is an interesting topic for future studies.

We now turn to the properties of eigenvectors (Fig.~\ref{eigvecs}). 
As noted in the introduction, the crucial feature of eigenvectors in the non-Hermitian case is that they are not mutually orthogonal. Thus, although the set of $d$ eigenvectors of a $d$-dimensional Hilbert space generically spans it, the eigenvectors ``effectively'' span a much lower-dimensional space. 
The effective dimension of this space can be estimated by looking at the singular value decomposition of the matrix $P$ that implements the transformation from the computational basis to the eigenbasis. 
Although $P$ is nominally full-rank, one can approximate it to high accuracy with a low-rank matrix $P'$, by discarding its smaller singular values. An estimate of the appreciable-sized singular values of $P$ is the inverse participation ratio of the (normalized) list of singular values of $P$; this also provides an estimate of the dimension of the space ``effectively'' spanned by $P$ (i.e., expressible in terms of eigenvectors without anomalously large coefficients). This effective dimension turns out to scale as $2^{-\alpha(\gamma)L}$ where $\alpha$ decreases from unity (which is the exact value in the Hermitian limit) to approximately $0.64$ at the transition. So \emph{almost all} vectors in the Hilbert space are ``difficult'' to express in the eigenbasis, viz. they can be expressed in this basis but with large coefficients that almost cancel out. 

It follows that a generic state, expanded in terms of the eigenbasis, has very large weight on a small number of eigenvectors. We now address what these eigenvectors are like. This question is generally more subtle in the non-Hermitian case, as one must distinguish between left and right eigenvectors; a full treatment is outside the scope of this work, but we can bypass some of the obvious issues by considering the entanglement properties of individual eigenvectors. These entanglement properties are often used as a metric of whether eigenstates are ``thermal''. We show how eigenstate entanglement evolves with $\gamma$ in the lower panels of Fig.~\ref{eigvecs}: for small $\gamma$ one has the standard ``thermal'' curve with low entanglement at the spectral edges and high entanglement in the middle of the spectrum. However, for $\gamma$ near the transition, the entanglement structure changes drastically: eigenvectors at similar eigenvalues have drastically different entanglement entropies, signaling a breakdown of eigenstate thermalization. Along the solvable line $\gamma = 1$, eigenvectors are manifestly not thermal, since generic eigenstates feature some fraction of spins pointing deterministically in the $|+z\rangle$ direction.

%An interesting topic for future work would be to study the correlations between eigenvectors, following Chalker and Mehlig**. We have not explored this topic in depth, and it does not seem to have been studied in the context of PT-symmetric quantum mechanics. 

\section{Discussion}

The central result of this work is that the entanglement transition in non-Hermitian quantum mechanics can be understood as a breaking of PT-symmetry in the many-body spectrum. This PT-symmetry breaking transition appears to be fundamentally different from the transition in random circuits---for example, it can happen even for a two-level system---but the overall phase diagrams are similar. In both cases, there is a phase in which an initially mixed density matrix stays mixed (and thus the system is able to retain quantum information) and an initial product state acquires volume law entanglement, as well as a phase where an initially mixed state purifies and a product state only develops area-law entanglement. To the resolution we have, the entanglement and purification transitions happen at the same point in this model, as they do in random circuits: the spectra evolve smoothly until they hit the first exceptional point. The existence of a fixed Hamiltonian and time-evolution operator give us some new tools to study this phenomenon, such as the ability to go into a diagonal ensemble and to study spectral statistics and eigenstate entanglement. Our results show how eigenstate entanglement both decreases on average and becomes much more heterogeneous across eigenstates as one approaches the exceptional point. The slow growth of entanglement in the volume law phase is an interesting feature of the dynamics of this model, for which we lack a simple intuition. 

The flow of quantum information under non-Hermitian time evolution is an interesting question raised by our work. In the example of the two-level system near its exceptional point, we saw that a generic vector expressed in the eigenbasis has very large coefficients, which are needed to capture the behavior along the ``missing'' dimensions. Under time evolution, this initial vector dephases and is rotated into the space spanned by the eigenvectors. However, naively it seems that the quantum information carried by the late-time state (i.e., its \emph{coefficients}) is determined precisely by these coefficients along the ``missing'' dimensions. The implications of this observation for encoding and retrieving quantum information remain to be understood. 

Finally, we remark that from a certain natural perspective, it is surprising that PT symmetry should survive to a nonzero value of $\gamma$, as empirically it seems to. If one thinks of the unperturbed Ising model as effectively a random matrix, a small anti-Hermitian perturbation would have matrix elements that are much larger than the level spacing between adjacent energy eigenvalues. One might expect that these would immediately give rise to exceptional points and destabilize the mixed phase. Evidently this does not happen, and, indeed, the eigenvalues exhibit a number of avoided level crossings before two of them eventually collide. It is unclear whether this is a general property of local PT symmetric Hamiltonians, or something more specific to the model we have studied here. %The resolution to this puzzle is not clear at present.

\emph{Note added}.---While this work was being completed, Ref.~\cite{biella2020many} was posted. Both works discuss non-Hermitian phase transitions but otherwise the models and results do not overlap. 

\acknowledgements{We thank Chris Hooley, David Huse, Vedika Khemani, and Vadim Oganesyan for helpful discussions and collaborations on related topics. M.J.G also thanks Yidan Wang and Alexey Gorshkov for helpful discussions.  This work was supported by the National Science Foundation under NSF Grant No. DMR-1653271.}

\bibliography{measurements,randlind}

% Adding sup mat
%\bigskip
%\onecolumngrid
%\newpage
%\includepdf[pages=1]{sup_mat.pdf}
%\newpage
%\includepdf[pages=2]{sup_mat.pdf}
%\newpage
%\includepdf[pages=3]{sup_mat.pdf}
%\newpage
%\includepdf[pages=4]{sup_mat.pdf}

\end{document}